\documentclass[aps, prr, reprint, longbibliography]{revtex4-2}
\usepackage{color}
\usepackage{mathtools}
\usepackage{lipsum}
\usepackage{hyperref}
\usepackage{graphicx}% Ilclude figure files
\usepackage{bm}% bold math
\usepackage{float}
\usepackage{epstopdf}
\usepackage{textcomp}
\usepackage[shortlabels]{enumitem}
\usepackage{verbatim}

\def\be{\begin{equation}}
\def\ee{\end{equation}}
\def\bea{\begin{eqnarray}}
\def\eea{\end{eqnarray}}

\def\nn{\nonumber}
\def\f{\frac}

\def\l{\left(}
\def\r{\right)}
\def\ls{\left[}
\def\rs{\right]}

\def\mr{\mathrm}
\def\refn{Eq.\,\ref}

\newcommand{\cmcb}{\affiliation{Centre for Mechanochemical Cell Biology and Division of Biomedical Sciences, Warwick Medical School, University of Warwick, Coventry CV4 7AL, United Kingdom}}

\newcommand{\mpipks}{\affiliation{ Max Planck Institute for the Physics of Complex Systems, Dresden, 01187, Germany}}

\newcommand{\csbd}{\affiliation{Center for Systems Biology, 01307 Dresden , Germany}}

\newcommand{\pol}{\affiliation{Cluster of Excellence Physics of Life, Technische Universit¨at Dresden, 01397 Dresden, Germany}}

\begin{document}

\title{Electrohydraulic Fields Generated by Active Transport at Tissue Interfaces}

\author{Amit Singh Vishen}
\email{amit.singh-vishen@warwick.ac.uk}
\cmcb
\mpipks

\author{Ahandeep Manna}
\mpipks

\author{Frank J\"ulicher }
\email{julicher@pks.mpg.de}
\mpipks
\csbd
\pol

\date{\today}

\begin{abstract}

Living cells and tissues can generate complex patterns of electric fields and fluid flows which can play important role in physiology. Both, fields and flows are rooted in ion transport across biological interfaces — cell membranes and epithelial cell layers. Here we develop a unified electrohydraulic framework that combines electric fields, osmotic pressures, and fluid flows, emphasising their couplings. 
We consider an active, permeable interface that drives electrohydraulic fields in the surrounding bulk.  We show that spatially heterogeneous ion transport acts as a distributed current source, generating long-range electric fields, osmotic gradients, and fluid flows. Using this framework, we show that patterns of ion pumping at cell and tissue boundaries can simultaneously produce large-scale electric fields and fluid flows due to electrohydraulic coupling. A key insight is that an external electric field and an internal dipolar pumping pattern can be physically equivalent-they can generate the same pattern of ion current and fluid flows. The induced dipolar osmotic pressure can drive self-propulsion through bulk osmotic coupling, with a mobility determined by interfacial permeability and system size, a mechanism distinct from classical electrophoresis or electro-osmosis. We further show that for strong fields a new effect emerges. Nonlinear coupling can lead to isotropic swelling of a hollow ball of cells. This can explain recent experiments on epithelial organoids. Finally, we show that feedback between ion transport and resulting electric fields can drive spontaneous symmetry breaking, generating dipolar or multipolar fields and patterns. Our work highlights the importance of electrohydraulic coupling in the emergence in currents and fields in the biological systems. 

\end{abstract}

 \maketitle

\section{Introduction}

Electric fields and ionic currents play a fundamental role in cells and tissues. While their function is well established in electrically excitable systems such as neurons and cardiac cells, they have increasingly been shown to play an important role in non-excitable contexts too, from embryonic development \cite{Levin2021, Nunes2025, Borgens1977, Adams2007, McCaig2005} and regeneration \cite{Liu2026} to collective cell migration \cite{Zhao2006, Shim2021, Kennard2020, Ferreira2025}. Equally, osmotic and hydraulic forces are now recognised as important drivers of morphogenesis, particularly in processes such as lumen formation and volume regulation \cite{Chan2019, Mosaliganti2019, Dagher2024, Duclut2021, Chugh2022, Wu2025}.
Although electric and hydraulic effects are frequently studied in isolation, ion transport provides a direct physical link between them. Ionic fluxes generate electric potentials, establish osmotic gradients, and drive water flow across biological boundaries \cite{Torres2021}; conversely, electric fields drive ion transport, creating a natural feedback between the two. Motivated by this coupling, we study electric and hydraulic effects together within the framework known as electrohydraulics \cite{Duclut2019, Torres2021}.

Electrohydraulic coupling has been developed within nonequilibrium transport theory to describe ionic and fluid fluxes across biological interfaces \cite{Stroka2014, Torres2021, Duclut2019, Duclut2021, Popovic2022}. Over the past decades, substantial theoretical progress has been made through complementary frameworks, including one-dimensional electrodiffusion models \cite{Li2015}, electrokinetic theories of charged and impermeable surfaces \cite{Squires2004}, and leaky-dielectric or capacitive descriptions with prescribed boundary currents \cite{Leonetti1997}. These approaches have clarified key mechanisms such as osmotic flows driven by solute fluxes and electrophoretic redistribution of charged membrane components under tangential electric fields \cite{Leonetti1997}. 

Building on this progress, an important open direction is to understand how ion transport, electric potentials, and hydrodynamic flows interact when treated self-consistently as spatially extended fields generated by active, permeable interfaces such as cell membrane and epithelial cell layers. In particular, the large-scale electric, osmotic, and flow fields produced by distributed ion transport at cellular and tissue scales remain less explored.

Beyond endogenous bioelectric phenomena, externally applied electric fields are widely used to probe and control cellular behaviour. Most studies focus on galvanotaxis, where applied fields bias transport, migration, or polarity along the field direction, producing anisotropic responses \cite{Zhao2006, Shim2021,Belliveau2025}. Recently, however, a qualitatively different effect has been reported: epithelial organoids subjected to a uniform electric field undergo isotropic lumen swelling, with the volume increase scaling with field strength \cite{Shim2024}. 

Here we study how electrohydraulical fields emerge from patterns of active ion transport. We focus on interplay between endogenous and external electric field, dynamic instabilities, pattern formation, and non-linear effects. Overall, we present a general framework for the coupling between electric and hydraulic processes in cells and tissues.

The paper is organised as follows. 
We first develop in Sec. II an electrohydraulic theory taking into account active ion transport at an interface.  
We show in Sec. III that spatial variations in active ion transport generate bulk electrohydraulic fields, in particular, electric potentials and osmotic gradients. 
In Sec. IV we discuss the response of spherical cavity with isotropic ion transport in an external electric field. We show the equivalence between the electrohydraulic fields generated by an external electric field and those of a dipolar pump pattern. In addition, we show that external fields can lead to isotropic inflation via non-linear coupling between electric field and ion transport. 
Finally, in Sec. V, we analyse how classical feedback mechanisms, such as voltage-dependent transport \cite{Fromherz199} and electrophoretic redistribution \cite{Jaffe1977, Poo1977}, operate within a spatially extended electrohydraulic field theory.

\section{Electrohydraulic Theory}

\begin{figure*}
    \centering
    \includegraphics[width= \linewidth]{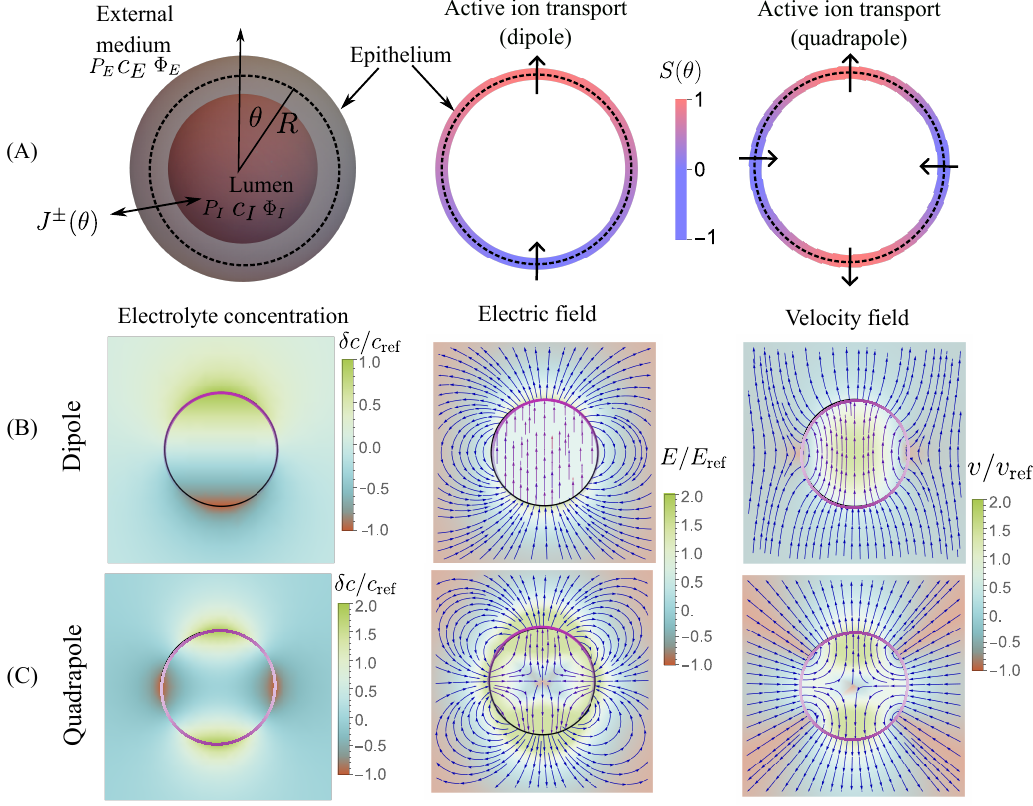}
    \caption{
\textbf{Active ion transport pattern generates bulk electrohydraulic fields.}
(A) From left to right: schematic of a spherical, permeable epithelial interface defining the geometry and spherical coordinates $(R,\theta)$; axisymmetric representations of dipolar and quadrupolar active ion transport patterns $S(\theta)$ on the surface.
(B) Bulk electrohydraulic response to a dipolar ($l=1,m=0$) active ion transport pattern.
From left to right: salt concentration perturbation $\delta c$, electric field $\mathbf E=-\nabla\Phi$, and fluid velocity $\mathbf v$ inside and outside the sphere.
(C) Response to a quadrupolar ($l=2,m=0$) active ion transport pattern. The angular structure of the bulk fields is fixed by the multipolar structure of surface transport, while their amplitude is controlled by interfacial permeability, size,  and diffusion as given in Eqs.~\eqref{eq:sol_c_main}-\eqref{eq:soln-ext}.
}

    \label{fig1}
\end{figure*}

We consider a minimal electrohydraulic system consisting of a closed, permeable interface separating two fluid compartments. The interface mediates ion and water transport between an interior and an exterior medium. For analytical tractability, the interface is taken to be spherical, enclosing an internal fluid domain and surrounded by an external fluid medium. Ion and water transport across the interface are incorporated at the coarse-grained level by effective surface permeabilities (Fig.~\ref{fig1}(A)).\\

\noindent{\bf Ion dynamics:} The interior and exterior fluids are treated as electrolytes composed of two ionic species, $c^+$ and $c^-$, with unit positive and negative charge, respectively. The dynamics of the ions are governed by the Nernst--Planck equation
\begin{equation}
\label{eq:iondynamics}
\partial_t c^\pm + \nabla \cdot (c^\pm \mathbf{v}) =
D^\pm \nabla \cdot \left( \nabla c^\pm \pm \beta c^\pm \nabla \Phi \right),
\end{equation}
where $\mathbf{v}$ is the fluid velocity, $D^\pm$ are the diffusion constants, $\beta = e/k_B T$, and $\Phi$ is the electrical potential.

At length scales much larger than the Debye length, the electrolyte is electroneutral, so we impose the constraint $c = c^+ = c^-$. The electric potential can then be understood as a Lagrange multiplier needed to enforce this constraint. Using electroneutrality in Eq.~(\ref{eq:iondynamics}) and subtracting the equations for positive and negative ions yields an equation
for the electric potential
\begin{equation}
\label{eq:potential}
\beta \nabla \cdot (c \nabla \Phi) =
\frac{D^- - D^+}{D^+ + D^-}\,\nabla^2 c .
\end{equation}
Adding the two equations and using Eq.~(\ref{eq:potential}) gives the evolution equation for the ion concentration,
\begin{equation}
\label{eq:iondynamics1}
\partial_t c + \nabla \cdot (c \mathbf{v}) =
\frac{2 D^+ D^-}{D^+ + D^-} \nabla^2 c .
\end{equation}
From Eq.~\eqref{eq:potential} we see that the electric potential is slaved to concentration gradients through the diffusion potential, which arises when the ionic diffusivities differ ($D^+ \neq D^-$) \cite{Anderson1989}. The average concentration evolves independent of electric potential with an effective diffusion constant given by the harmonic mean of the individual diffusivities.

The governing equations are solved in the interior and exterior domains subject to boundary conditions at the interface and in the far field. Far away from the sphere we impose $c(r) = c_E$ and $\Phi(r) = 0$, as $r \to \infty$. The two compartments, outside and inside the sphere, are coupled through boundary conditions at the spherical interface.

Within linear nonequilibrium thermodynamics, ionic and fluid fluxes are linear functions of the corresponding thermodynamic forces through a constant Onsager matrix \cite{Katchalsky1965,deGroot2013}. We work in a decoupled limit in which cross-coupling terms are negligible, while retaining the nonlinear dependence of ionic transport on the electrochemical potential difference. The radial ionic flux across the interface is therefore written as
\begin{equation}
\label{eq:ion-bc}
J_r^\pm =
\Lambda^\pm \left( c_I - c_E e^{\mp \beta \Delta \Phi} \right) + S^\pm ,
\end{equation}
where $\Lambda^\pm$ are passive permeabilities, $S^\pm$ represent active ion transport, and $\Delta \Phi = \Phi_I - \Phi_E$ is the electrochemical potential difference across the interface. Subscripts $I$ and $E$ denote interior and exterior quantities, respectively. For an epithelium, this effective surface flux combines transcellular and paracellular transport at the coarse-grained level and recovers electrochemical equilibrium in steady state in the absence of active pumping.

The surface flux $J_r^\pm$ enters the bulk electrodiffusion problem as a boundary condition at the interface. Specifically, continuity of ionic flux requires that the normal component of the bulk Nernst--Planck flux matches the interfacial flux,
\begin{equation}
\label{eq:flux-continuity}
\hat{\mathbf{n}}\cdot
\left[
c^\pm ({\bf v} - \dot R ~{\bf \hat n}) - D^\pm \left( \nabla c^\pm \pm \beta c^\pm \nabla \Phi \right)
\right]_{r=R}
=
J_r^\pm ,
\end{equation}
where $\hat{\mathbf{n}}$ is the outward unit normal to the interface, $\dot R$ is the rate of change of radius of the sphere.\\

\noindent{\bf Fluid flow:} In the low Reynolds number regime, fluid flow inside and outside the sphere is governed by the Stokes equation,
\begin{equation}
\label{eq:fludidynamics}
\eta \nabla^2 \mathbf{v} - \nabla P = 0 ,
\end{equation}
together with the incompressibility condition $\nabla \cdot \mathbf{v} = 0$, where $\eta$ is the viscosity and $P$ is the hydrodynamic pressure \cite{Happel2012}.

At the interface, the normal component of the velocity relative to the moving boundary is determined by the transepithelial fluid flux,
\begin{equation}
\label{eq:fluid-bc}
v_r(R) - \dot{R} =
\Lambda_w \left[ \Delta P - \Delta \Pi  \right],
\end{equation}
where $\Lambda_w$ is the hydraulic permeability of the interface, $\Delta P \equiv (P_I - P_E)$ and $\Delta \Pi \equiv (\Pi_I - \Pi_E) $ are the hydrostatic and osmotic pressure differences, respectively. No-slip boundary conditions are imposed for the tangential velocity components at the interface, so that $v_\theta(R) = v_\phi(R) = 0$.

Eqs.~(\ref{eq:potential})--(\ref{eq:fluid-bc}) provide a closed and thermodynamically consistent description of coupled ion transport and fluid flow in the interior and exterior domains within an electroneutral, coarse-grained framework. This formulation applies at scales much larger than the Debye length and ion transporter spacing, where pumps and channels can be treated as a continuous transport density and bulk electrodiffusion maintains electroneutrality. The electric potential is set by transport constraints rather than by resolving microscopic charge separation with Poisson’s equation. At microscopic scales, however, discrete membrane structures and local charge distributions require solving Poisson’s equation. The coarse-grained approach here is suitable for cellular and tissue scales, where fine details are averaged out and interfacial currents define boundary behaviour.

\section{Active interface transport drives bulk electrohydraulic fields}

We now analyse how spatially heterogeneous ion transport across an active, permeable interface generates bulk electrohydraulic fields. Our goal is to determine how nonuniform surface activity gives rise to electric fields, osmotic gradients, and fluid flows in the surrounding medium, and to quantify the magnitude of these effects. We proceed by constructing the isotropic steady state corresponding to uniform surface transport and then analysing the linear response to weakly anisotropic surface activity.

The characteristic diffusion time is $\tau \sim R^2/D$: for $R \approx 10$--$100~\mu\mathrm{m}$ and $D \approx 2000~\mu\mathrm{m}^2/\mathrm{s}$, one finds $\tau \approx 0.05$--$5~\mathrm{s}$. For physiological velocities $v \lesssim 1~\mu\mathrm{m}/\mathrm{s}$, the P\'eclet number satisfies $vR/D \lesssim 0.1$. Thus, on minute timescales and for such velocities, we neglect advection and work in the steady-state limit.
In this limit, Eq.~\eqref{eq:potential}  and Eq.~\eqref{eq:iondynamics1} reduce to
\begin{align}
\nabla^2 c &= 0,
\label{eq:laplace_c}
\\
\nabla\cdot\!\left(c\,\nabla\Phi\right) &= 0,
\label{eq:laplace_phi}
\end{align}
in both the interior and exterior domains. 

In the same limit, the boundary condition in Eq. \ref{eq:flux-continuity} reduces to 
\begin{equation}
\label{eq:flux-continuity1}
\hat{\mathbf{n}}\cdot
\left[
 - D^\pm \left( \nabla c^\pm \pm \beta c^\pm \nabla \Phi \right)
\right]_{r=R}
=
J_r^\pm .
\end{equation}
These equations, together with flux continuity at the interface, fully determine the bulk electrohydraulic fields generated by active surface transport. For simplicity, we take equal ionic diffusivities, $D^+=D^-=D$; while unequal diffusivities generate diffusion potentials during transient dynamics, they do not alter the steady-state bulk field structure considered here.

For uniformly distributed pumps and channels, spherical symmetry implies that only the isotropic component of the transport enters the steady state, which fixes the interior and exterior concentrations and the steady trans-epithelial potential difference (see Appendix~A).

We now introduce patterns of ion transport by varying the  surface transport parameters along the interface. Small deviations from the isotropic steady state are treated perturbatively.  For small perturbations, the concentration and electric potential satisfy Laplace equations in the bulk (Eqs.~\eqref{eq:laplace_c} and \eqref{eq:laplace_phi}). The interfacial ion flux, given by Eqs.~\eqref{eq:ion-bc} and \eqref{eq:flux-continuity1}, provides boundary conditions through continuity of ionic flux across the interface. Together with electroneutrality, these relations yield linear boundary conditions that couple perturbations in surface transport activity to the bulk concentration and electric potential fields. The resulting solutions can be decomposed into independent angular modes using the standard scalar spherical harmonics ($Y_{lm}(\theta,\phi)$) as $\delta c^{i} =\sum_{l=0}^{\infty}\sum_{m=-l}^{l}\delta c_{lm}^{i}(r)Y_{lm}$ and $\delta \Phi^{i}=\sum_{l=0}^{\infty}\sum_{m=-l}^{l}\delta \Phi_{lm}^{i}(r)Y_{lm}$, where $i\in\{I,E\}$, $l$ is the angular mode number, and $m$ is the azimuthal index. The full analytical solution is presented in Appendix~B; here we focus on the physical implications.

A key control parameter is the dimensionless ratio $D/(R\Lambda^\pm)$, which compares the timescale for bulk diffusive relaxation to that for interfacial ion exchange. 
For typical cell and tissues parameter values:ionic diffusivities $D\sim10^3~\mu\mathrm{m}^2/\mathrm{s}$, radii $R\sim 10$--$100~\mu\mathrm{m}$, and passive permeabilities $\Lambda^\pm\sim10^{-3}$--$1~\mu\mathrm{m/s}$ \cite{Torres2021}, the dimensionless ratio $D/(R\Lambda^\pm)$ ranges from $\mathcal{O}(10)$ to $\mathcal{O}(10^5)$. Thus, bulk diffusion is generically fast compared to interfacial exchange in biological tissues, so the active transport behaves as a \emph{distributed constant current source} coupled to the bulk.

We further take equal passive permeabilities for cations and anions, $\Lambda^+=\Lambda^-=\Lambda$; relaxing this assumption rescales the response amplitudes but does not modify the physical structure of the solutions. 
In this limit, for $l\neq 0$, the interior concentration and electric-potential perturbations take the form 
\begin{align}
\delta c_{lm}^I
&=
- \frac{R\Lambda}{lD}\left(s_{lm}^+ + s_{lm}^-\right),
\label{eq:sol_c_main}
\\
\beta \bar c_I \,\delta\Phi_{lm}^I
&=
- \frac{R\Lambda}{lD}\left(s_{lm}^+ - s_{lm}^-\right),
\label{eq:sol_phi_main}
\end{align}
where $s_{lm}^\pm$ denotes the angular components of the surface transport perturbation. 
Symmetric modulation of cation and anion transport generates osmotic concentration gradients, while antisymmetric modulation generates electric potentials. The ratio $D/(R\Lambda)$ controls the amplitude of the response, the multipolar structure of the bulk fields is fixed by the surface transport pattern. 
The exterior concentration and potential perturbations follows from the continuity of ionic flux at the interface as given by Eq.~\ref{eq:flux-continuity1}. For $l \neq 0$, one obtains
\begin{equation}
\label{eq:soln-ext}
c_{lm}^E = -\frac{l}{l+1}\, c_{lm}^I, 
\quad
 \Phi_{lm}^E = -\frac{l}{l+1}\, \frac{\bar c_I}{\bar c_E} \Phi_{lm}^I.
\end{equation}
\\

\noindent\textbf{Fluid flow and self-propulsion.}
The concentration gradients generated by anisotropic ion transport drive fluid motion through osmotic pressure differences across the interface. Solving the Stokes equations in the interior and exterior domains with the  boundary conditions, given by Eq.~\eqref{eq:fludidynamics} and Eq.~\eqref{eq:fluid-bc} \cite{Kim2013}, and substituting the concentration yields a normal surface velocity proportional to the osmotic-pressure harmonics. For each angular mode,
\begin{equation}
v^r_{lm}(R)
=
-\frac{\Lambda_w R}{R+\Lambda_w\eta\,\mathcal{K}_l}\,
\Delta\Pi_{lm},
\label{eq:vr_main}
\end{equation}
where $\Delta\Pi_{lm}$ is the corresponding osmotic-pressure difference and $\mathcal{K}_l$ is a dimensionless number that is a function of $l$,   determined by the Stokes solution (see Appendix B). 
A notable consequence of this electrohydraulic coupling is \emph{self-propulsion}. For dipolar ($l=1$) components of the surface transport, the induced osmotic pressure gradients generate a net fluid flux that produces a finite propulsion velocity of the interface, where the velocity is given by
\begin{equation}
v_0
=
\sqrt{\frac{3}{4\pi}}\;
\frac{\Lambda_w R}{2\left(R+10\,\Lambda_w\eta\right)}\,
\Delta\Pi_{10}.
\label{eq:v0_main}
\end{equation}
This self-propulsion driven by asymmetric solute fluxes has been analysed previously in the context of osmotic and electro-osmotic motion of permeable interfaces for imposed ion pattern \cite{Leonetti1995, Lammert1996}. More broadly, this mechanism complements other self-propulsion strategies of active particles, including electrophoresis, diffusiophoresis, and surface-slip-driven motion, by relying on osmotic pressure gradients generated through interfacial transport \cite{Anderson1989}.
Figures~\ref{fig1}(B--C) show the concentration, electric field, and flow fields generated by dipolar ($l=1$) and quadrupolar ($l=2$) components of the surface transport, which act as multipolar current sources at the interface. 

We can also obtain the hydrostatic pressure difference associated with each spherical harmonic mode. Substituting the velocity back into the expression for the pressure jump across the surface gives
\begin{equation}
\Delta P_{lm}
=
\frac{\Lambda_w\eta\,\mathcal{K}_l}
{R+\Lambda_w\eta\,\mathcal{K}_l}
\Delta\Pi_{lm}.
\label{eq:DeltaP}
\end{equation}
For the non-isotropic modes ($l\neq0$), the hydrostatic pressure difference is thus different than the osmotic pressure difference at steady state. In contrast, for the isotropic mode ($l=0$), the hydrostatic and osmotic pressure differences are equal at steady state.

We now estimate the magnitude of the electrohydraulic fields generated by anisotropic transport at the surface. For the dipolar ($l=1$) mode, taking a moderate surface-activity modulation $s_{10}/\bar c_I\sim0.1$ and parameters in the current--source regime $D/(R\Lambda)\gg1$, Eq.~\eqref{eq:sol_c_main} gives a concentration variation suppressed by the factor $R\Lambda/D$. For $D/(R\Lambda)\sim10$, this corresponds to $\Delta c/\bar c_I\sim10^{-2}$. Taking $\bar c_I\sim150~\mathrm{mM}$, this corresponds to $\Delta c\sim1.5~\mathrm{mM}$. Using  $\Delta\Pi_{10}\sim2k_BT\Delta c_{10}$, gives $\Delta\Pi_{10}\sim7.5~\mathrm{kPa}$ at room temperature. Thus, even a small concentration variation can generate a substantial osmotic pressure difference.

From Eq.~\eqref{eq:sol_phi_main}, we obtain $\beta\Phi\sim10^{-2}$, corresponding to an electric potential difference $\Phi\sim0.25~\mathrm{mV}$ at room temperature. The corresponding electric field is of order $E\sim\Phi/R$, giving $E\sim 0.025$--$0.25~\mathrm{V/cm}$ for $R\sim10$--$100~\mu\mathrm{m}$.

For typical parameter values, $\Lambda_w\sim10^{-7}\,\mu\mathrm{m\,Pa^{-1}\,s^{-1}}$, $\eta\sim10^{-3}\,\mathrm{Pa\,s}$, and $R\sim100\,\mu\mathrm{m}$, we find $\Lambda_w\eta/R\sim10^{-12}\ll1$, giving $\Delta P_{lm}/\Delta\Pi_{lm}\sim10^{-12}\mathcal{K}_l$ for $l\neq0$. For the dipolar mode ($l=1,m=0$), $\mathcal{K}_1=10$, and using the osmotic pressure scale estimated above gives $\Delta P_{10}\sim10^{-7}$~Pa. Thus, the hydrostatic pressure associated with the non-isotropic modes is small compared with the corresponding osmotic pressure difference. Consequently, the hydrostatic pressure variations associated with the anisotropic modes do not produce significant deformation of the  surface, given that the elastic moduli of cells and epithelial tissues are in the $10$--$100$~Pa range. In the biologically relevant limit $\Lambda_w\eta/R\ll1$, the non-isotropic osmotic pressure differences therefore primarily drive interfacial fluid flow rather than hydrostatic pressure buildup. Using $R\sim10~\mu\mathrm{m}$, representative of cellular length scales, Eq.~\eqref{eq:v0_main} gives a self-propulsion velocity $v_0\sim0.1~\mu\mathrm{m/min}$, comparable to measured cell migration speeds \cite{Trepat2009}.

\section{Electrohydraulic response to external electric field}

\begin{figure*}
    \centering
    \includegraphics[width= 0.9\linewidth]{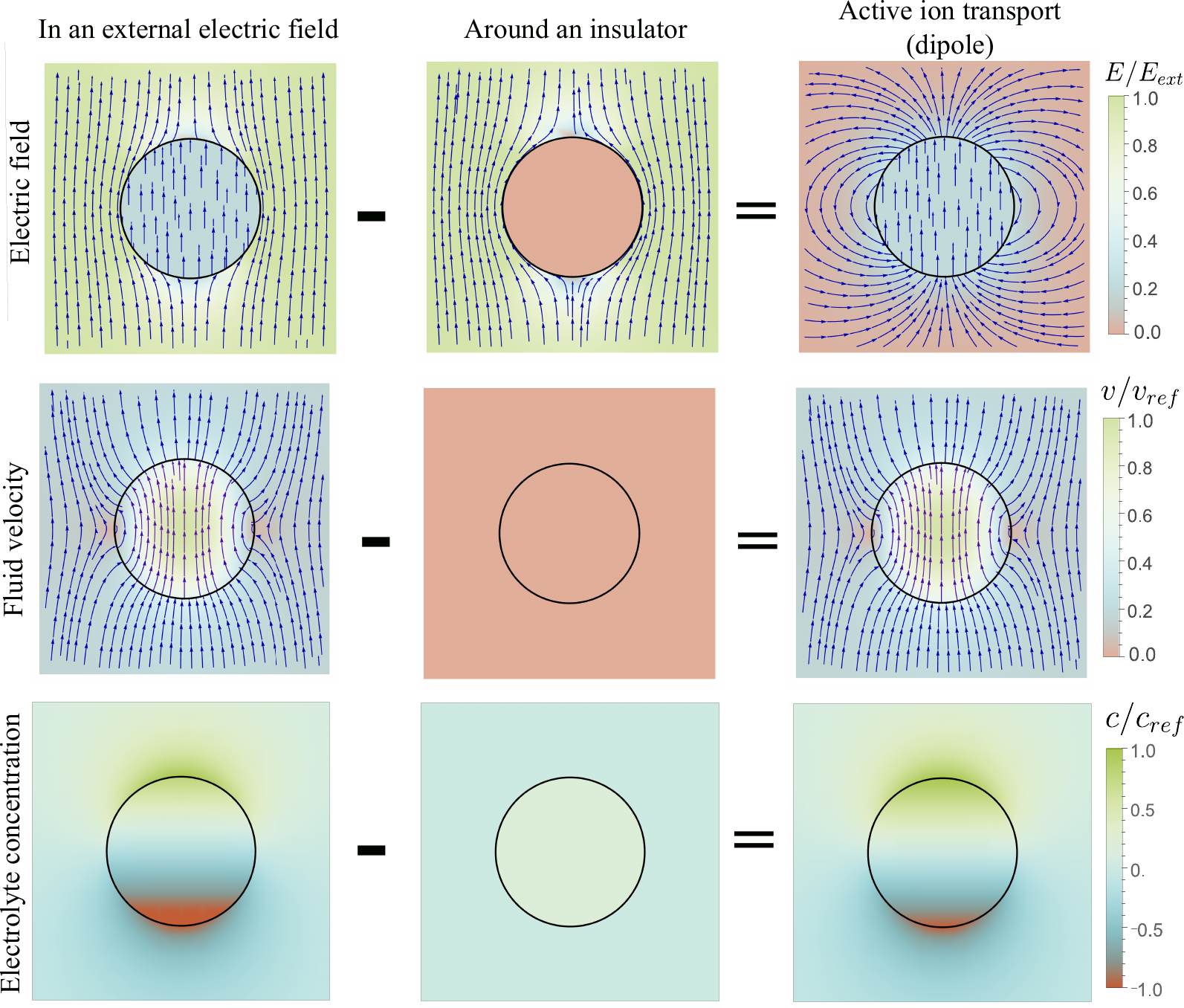}
\caption{
\textbf{Electrohydraulic equivalence between external electric fields and active ion transport.}
Left column: first-order ($\mathcal{O}(\epsilon)$) electrohydraulic fields obtained from the coupled ion-transport and fluid equations.
Middle column: response of a perfectly insulating epithelial surface ($\Lambda^\pm = 0$), for which the electric field is purely tangential, ion concentrations remain uniform, and no fluid flow is generated.
Right column: difference between the two cases, isolating the bulk electrohydraulic response generated by finite epithelial permeability.
This difference corresponds to an effective dipolar ($l=1$) surface current source induced by the external field, producing long-range electric, osmotic, and hydrodynamic fields.
}
    \label{fig2}
\end{figure*}

Having examined polarisation driven by imposed ion transport asymmetries, we now consider a distinct source of asymmetry: a uniform external electric field. Unlike internally generated fields, an external field provides a controlled perturbation that explicitly breaks spatial symmetry. 

A recent experiment exploited this very type of controlled forcing, demonstrating that applying a constant electric field to epithelial organoids induces isotropic lumen swelling \cite{Shim2024}. This isotropic response provides a clear experimental signature of nonlinear electrohydraulic transport. In what follows, we use our theoretical framework to analyse linear and nonlinear responds under uniform electric forcing, and to identify the mechanisms responsible for electrically induced volume expansion.

The governing equations for ionic transport contain two key nonlinearities that are central to the electrohydraulic response: (i) the exponential dependence of transmembrane ionic currents on the electric potential difference (see Eq.~\eqref{eq:ion-bc}), and (ii) the coupling between the electric potential and ion concentration in the steady-state bulk equations (see Eq.~\eqref{eq:laplace_phi}). As in the previous section, we analyse the system's linear response to a uniform externally applied electric field. To account for the experimentally observed behaviour, we then extend the analysis to include nonlinear effects by expanding to second order in the applied perturbation. 

We work in the weak–field regime \(\beta E_\mathrm{ext} R \ll 1\). For bookkeeping, we introduce a formal small parameter \(\epsilon \ll 1\) and write the uniform applied electric field directed along the \(z\)-axis as \(\mathbf{E}_\mathrm{ext} = \epsilon E_\mathrm{ext} \hat{\mathbf z}\), with \(\epsilon\) set to unity at the end of the calculation. The field may be expressed in terms of spherical vector harmonics as
\be
\mathbf{E}_\mathrm{ext}
= \epsilon E_\mathrm{ext} \sqrt{\frac{4\pi}{3}}
\left(
\mathbf Y_{10}
+ \mathbf\Psi_{10}
\right),
\ee
where $\mathbf{Y}_{10}$ and $\mathbf{\Psi}_{10}$ denote the radial and tangential vector spherical harmonics, respectively.
The corresponding external electrostatic potential is $\Phi_\mr{ext} = -\epsilon \,E_\mr{ext} \sqrt{\f{4\pi}{3}} r Y_{10},$ with the potential at the centre of the sphere chosen as the reference. We expand the electric potential and ionic concentrations to second order in $\epsilon$ as $c^{i} = c_0^i + \epsilon c_1^i + \epsilon^2 c_2^i + \mathcal{O}(\epsilon^3)$, $\Phi^i = \Phi_0^i + \epsilon (\Phi_1^i + \Phi_\mr{ext}) + \epsilon^2 \Phi_2^i + \mathcal{O}(\epsilon^3)$, where $i \in \{ I,E\}$ denotes the internal and external medium. We proceed by solving the governing equations, Eqs.~\eqref{eq:laplace_c}, \eqref{eq:laplace_phi},~\eqref{eq:ion-bc} and \eqref{eq:flux-continuity1}, order by order in $\epsilon$. 

At zeroth order there is no driving field and the concentrations $c^i_0$ and electric potentials $\Phi_0^i$ are constants in space both inside and outside the cavity. \\

\noindent\textbf{Electrohydraulic equivalence between external electric fields and active ion transport.} 
%\subsection{Linear response}
At first order, different spherical harmonic modes are decoupled. Since the external forcing contains only the $l=1,m=0$ component, the only non-vanishing first-order response occurs in the $l=1,m=0$ mode. The corresponding solution is identical to that obtained previously in the absence of an external field, but with an effective source term given by (see Appendix C)
\be 
s_{lm}^{\pm} = \pm \epsilon\, c_{0}^E e^{\mp\beta\Delta \Phi_0} \f{R\beta E_{ext}}{2}\delta_{10,lm}.
\label{eq:effective-s}
\ee 
As discussed above, this effective current source induces both an electric dipole and a fluid source dipole. In the simplified limit of equal ionic diffusivities and permeabilities, the interior concentration and electrostatic potential amplitudes for the $l=1,m=0$ mode are
\begin{align}
\delta c_{10}^I
&=
- \frac{R^2\Lambda \beta}{D} c_{0}^E  \sinh(\beta \Delta \Phi_0) E_\mathrm{ext},
\label{eq:sol_c_E}
\\
\beta \,\delta\Phi_{10}^I
&=
- \frac{R^2\Lambda \beta}{D} \frac{c_{0}^E}{c^I_0}  \cosh(\beta \Delta \Phi_0) E_\mathrm{ext}.
\label{eq:sol_phi_E}
\end{align}
The solution in the exterior domain follows directly from Eq.~\eqref{eq:soln-ext}.
The electric field–induced dipolar ($l=1$) osmotic gradient generates fluid flow and produces self-propulsion of the active interface, as given by Eq.~(\ref{eq:v0_main}). This mechanism is distinct from electrophoresis and electro-osmosis: it does not rely on surface charge, charged double layers, or electro-osmotic slip \cite{Anderson1989}. Instead, a uniform external electric field induces an asymmetry in ion transport across the permeable interface, generating a dipolar osmotic pressure that drives motion through bulk electrohydraulic coupling.
As a result, the propulsion velocity is linear in the applied field,
\begin{equation}
    \mathbf{v} = \xi\, \mathbf{E}_\mathrm{ext},
\end{equation}
with a mobility $\xi$ determined by active transport parameters, osmotic permeability, and system size. Unlike classical phoretic mobilities, which depend primarily on interfacial charge or slip properties, the coefficient $\xi$ reflects the coupling between field-induced transport asymmetry and osmotic pressure generation. The precise form of $\xi$ can be obtained by computing the osmotic pressure difference from Eq.~\eqref{eq:sol_c_E} and substituting it in Eq.~\eqref{eq:v0_main}.  We refer to this mechanism as field-induced electrohydraulic propulsion. 
To our knowledge, such field-induced propulsion mediated purely by electrohydraulic coupling in the electroneutral regime has not been previously identified.

Fig.~\ref{fig2} illustrates the electrohydraulic response of the spherical cavity to a uniform external electric field at first order in the perturbative expansion. The left column shows the ion concentration, electric field, and fluid flow obtained from the \(\mathcal{O}(\epsilon)\) solution of the coupled electrohydraulic equations. The middle column presents the corresponding response of a perfectly insulating epithelial surface (for which the electric field is purely tangential to the spherical boundary, the fluid velocity vanishes and the ion concentrations are uniform across space). The right column displays the difference between these two cases, isolating the \(l=1\) dipolar contribution arising from finite epithelial conductivity. This comparison shows that the externally applied electric field induces an electrohydraulic response equivalent to that generated by an effective dipolar pattern of ionic pumping at the epithelial surface. \\

\begin{figure}[t]
    \centering
    \includegraphics[width=\linewidth]{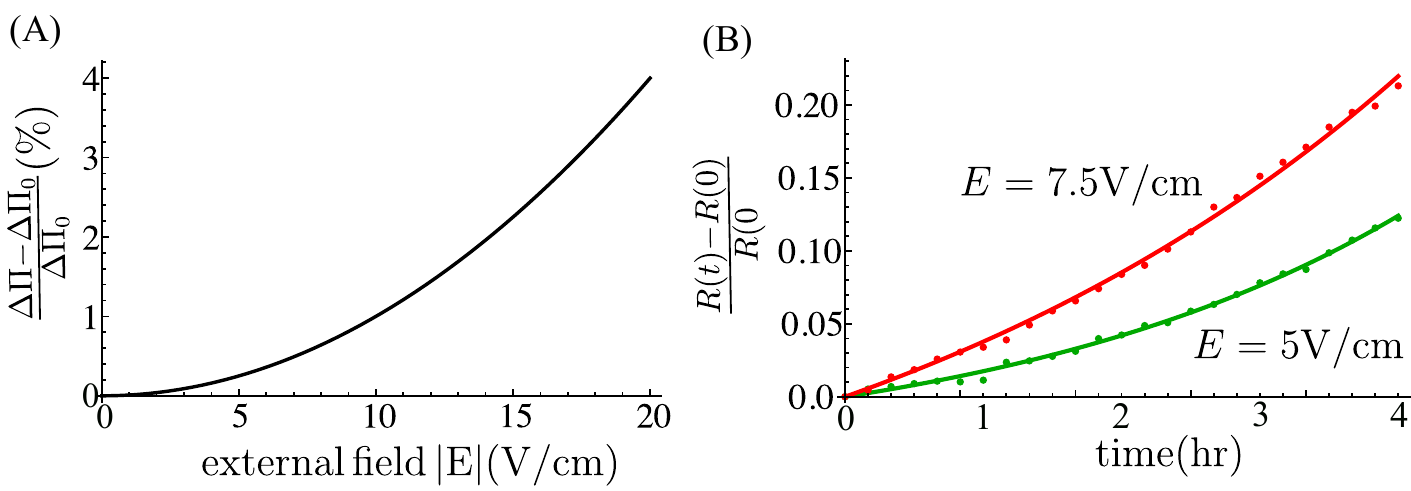}
    \caption{
\textbf{Isotropic inflation as a nonlinear response to an applied electric field.}
(A) Isotropic ($l=0$) change in osmotic pressure difference $\Delta\Pi - \Delta \Pi_0$ generated at second order in the applied electric field, showing the quadratic dependence on $|E|$ predicted by the nonlinear electrohydraulic theory (\refn{eq:pi-E}).
(B) Electric-field-induced swelling of epithelial organoids. Symbols show experimental measurements of the normalized lumen radius $R(t)/R(0)-1$ under uniform applied fields (data adapted from Fig.~1C of Ref.~\cite{Shim2024}); red: $E=7.5\,\mathrm{V/cm}$, green: $E=5\,\mathrm{V/cm}$. Solid lines show fits to the predicted growth law given by \refn{eq:dynamics-R}.
}
\label{fig3}
\end{figure}

\noindent\textbf{Isotropic inflation as a nonlinear response to an applied electric field.} At the second order in the perturbative expansion, both the $l=0$ and $l=2$ modes acquire a non-zero value. The $l=2$ mode does not lead to isotropic inflation of the organoid as the surface integral of the associated fluid flux vanishes, so that no net fluid accumulation can occur. 
In contrast, $l=0$ mode produces an isotropic response and leads to uniform swelling of the organoid. This swelling arises from an osmotic pressure difference established across the epithelium, which drives net fluid accumulation within the lumen. The resulting change in osmotic pressure is given by (see Appendix C)
\be
\label{eq:pi-E}
\frac{\Delta \Pi - \Delta \Pi_0}{\Delta \Pi_0} = \frac{1}{2}\cosh\l \beta \Delta \Phi_0\r\left(\frac{\beta R^2\Lambda}{D}\right)^2\left(\frac{c^E_0}{c^I_0} + \frac{1}{2}\right) E_\mathrm{ext}^2.
\ee 
Fig.~\ref{fig3}(A) show the increase in osmotic pressure different as a function of electric field.   Although this contribution is formally second order in the applied field, even a small osmotic imbalance is sufficient to induce significant inflation of the organoid.

We now use the osmotic pressure difference generated by the second-order electrohydraulic response derived in the previous section to determine the resulting lumen volume dynamics.

Previous studies of epithelial cysts indicate that water transport is typically the rate-limiting process, with a hydraulic permeability of order\(\Lambda_w \sim 10^{-7}\,\mu\mathrm{m}\,\mathrm{Pa}^{-1}\,\mathrm{s}^{-1}\) \cite{Chan2019,Shim2024}. In this regime, the dynamics can be accurately described by an effective evolution equation for the lumen volume, closely related to the models developed in Refs.~\cite{Shim2024,Chan2019}. 
The dynamics of spherical cavity of radius $R$ is given by
\begin{equation}
\frac{dR}{dt} = \Lambda_w \bigl( \Delta \Pi - \Delta P \bigr),
\end{equation} 
The hydrostatic pressure difference is given by Laplace’s law, \(\Delta P = 2\gamma/R\), where \(\gamma\) denotes the epithelial tension.

We now consider small deviations from an initial lumen radius \(R_0\), corresponding to the state before the electric field is applied, and write \(R(t) = R_0 + \delta R(t)\) with \(|\delta R| \ll R_0\). In general, the hydraulic permeability \(\Lambda_w\), the osmotic pressure difference \(\Delta \Pi\), and the hydrostatic pressure difference \(\Delta P\) may all depend on the instantaneous radius \(R\). Expanding the evolution equation to first order in \(\delta R\), and introducing the dimensionless radius perturbation \(\delta \tilde R \equiv \delta R / R_0\), the dynamics can be written in
the compact form
\begin{equation}
\label{eq:dynamics-R}
\partial_t \delta \tilde R
= b(E^2) + a(E^2)\,\delta \tilde R ,
\end{equation}
where the effective coefficients
$b = \Lambda_0 \bigl( \Delta \Pi_0 - \Delta P_0 \bigr)$, $a = \Lambda_0' \bigl( \Delta \Pi_0 - \Delta P_0 \bigr) + \Lambda_0 \bigl( \Delta \Pi_0' - \Delta P_0' \bigr)$, depend on the applied field strength through the osmotic pressure \(\Delta \Pi(E^2)\). Here we have defined \(\Lambda_0 = \Lambda_w(R_0)\), \(\Delta \Pi_0 = \Delta \Pi(R_0)\), \(\Delta P_0 = \Delta P(R_0)\), and primes denote derivatives with respect to \(R\) evaluated at \(R_0\).

We fit this linear evolution law to the experimental lumen swelling data reported in Ref.~\cite{Shim2024}.  Fig.~\ref{fig3}(B) shows the best fits for two electric field amplitudes, $E = 7.5\,\mathrm{V/cm}$ and $E = 5\,\mathrm{V/cm}$. Taking $\Lambda_0 \sim 10^{-7}\,\mu\mathrm{m}\,\mathrm{Pa}^{-1}\,\mathrm{s}^{-1}$, the fits yield $\Delta \Pi_0 - \Delta P_0 \approx \frac{b}{\Lambda_0} \approx 4000\,\mathrm{Pa}$ for  $E = 7.5\,\mathrm{V/cm},$ and $\Delta \Pi_0 - \Delta P_0 \approx 2000\,\mathrm{Pa}$ for $E = 5\,\mathrm{V/cm}$.
These pressure differences are sufficiently large that the contribution of the hydrostatic term $\Delta P_0 = 2\gamma/R_0$ can be neglected to leading order, allowing us to interpret them as estimates of the osmotic pressure jump $\Delta \Pi_0$. The ratio of the osmotic contributions extracted from the fits is close to the ratio $E_{\text{high}}^2 / E_{\text{low}}^2$, consistent with the theoretical prediction $\Delta \Pi \propto E^2$ shown in Eq.~\eqref{eq:pi-E}.

Our overall conclusion is similar in spirit to the theoretical interpretation of Ref.~\cite{Shim2024}: an increase in cyst volume under an applied electric field is driven by an increase in lumen osmolarity.
However, the underlying physical mechanism is fundamentally different.
In Ref. ~\cite{Shim2024}, electro-inflation is attributed to field-enhanced ion crowding in the diffuse double layer at the basal surface, producing polarized (anode–cathode) ionic distributions and increased near-surface ion concentrations that, together with epithelial ion transport, raise luminal osmolarity and drive osmotic water influx.
In contrast, our analysis operates at length scales much larger than the Debye length, developing a coarse-grained description of the tissue in which the epithelium is treated as a permeable and electrically active interface.

Finally, we emphasise that the observed shape change of the lumen is an electrohydraulic effect, arising from the coupling between ion transport, osmotic pressure, and fluid flow. This mechanism is fundamentally distinct from the electric-field–induced shape changes of lipid vesicles, which are typically governed by electro\-mechanical effects within the membrane itself, such as Maxwell stresses or field-induced changes in membrane tension, rather than by osmotic water transport \cite{Winterhalter1988,Vlahovska2009}.

\section{Spontaneous Electrohydraulic Polarisation}

\begin{figure}[t]
    \centering
    \includegraphics[width= \linewidth]{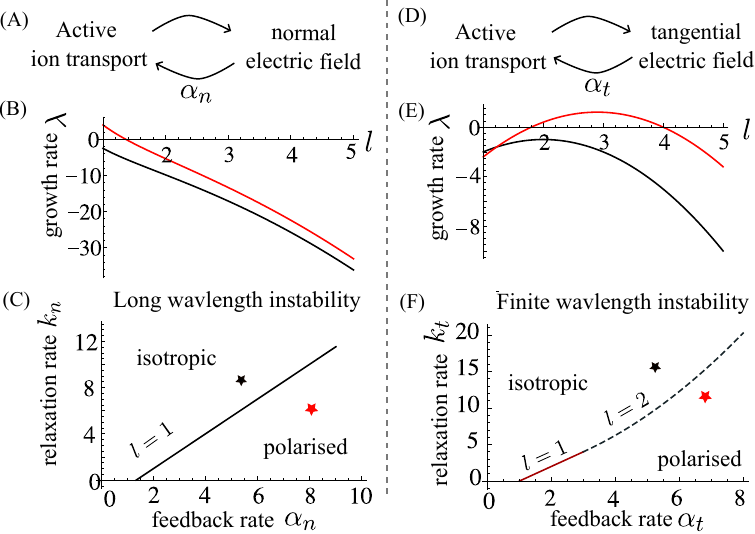}
\caption{
\textbf{Spontaneous electrohydraulic polarisation and mode selection.}
(A) Shows feedback between active ion transport and normal electric field.
(B) Linear stability spectrum for voltage-dependent feedback. Growth rate \(\lambda_l\) of spherical harmonic modes as a function of angular index \(l\). Instability first appears in the dipolar mode (\(l=1\)), corresponding to global polarisation. Black and red curves show stable and unstable growth rates, respectively, for \(k_n=8,\alpha_n=5\) (black) and \(k_n=6,\alpha_n=8\) (red) 
(C) Phase diagram in the plane of feedback strength and relaxation rate, showing the boundary between stable and unstable regimes for transport dependent on normal electric field. The stability boundary (solid black line) corresponds to the $l=1$ mode; since this mode is always the first to become unstable, crossing this boundary leads directly to long-wavelength ($l=1$) global polarisation. Black and the red stars show the parameter values used in fig (B).
(D) Shows feedback between active ion transport and tangential electric field. 
(E) Linear stability spectrum for electrophoretic feedback showing growth rate $\lambda_l$ versus $l$ for increasing feedback strength. In contrast to (A), instability can occur at finite $l$, selecting a characteristic angular scale. Black and red curves show stable and unstable growth rates, respectively, for $k_n = 10, \alpha_n = 5$  (black) and $k_n = 14, \alpha_n = 6.8 $ (red).
(F) Corresponding phase diagram for electrophoretic redistribution, showing the region of spontaneous polarisation in feedback–relaxation parameter space. Two stability boundaries are shown: the red solid line marks the onset of the $l=1$
mode and the black dashed line marks the onset of the $l=2$ mode. This shows that depending on the parameter value a finite-wavelength ($l=2$) patterned polarisation can be the dominant instability, in contrast to the long-wavelength instability selected by normal-field feedback. Black and the red stars show the parameter values used in fig (E).
}
    \label{fig4}
\end{figure}

Thus far, we have considered situations in which spatial asymmetry is imposed externally, either through prescribed patterns of ion transport or through applied electric fields. In living tissues, however, electrical polarity can also emerge spontaneously through feedback between electric fields and ion transport. Here we analyse how such feedback operates within the spatially extended electrohydraulic framework developed above.

Instabilities in electrodiffusive systems have been studied in various contexts, including pattern formation driven by bulk ion diffusion and electric fields \cite{Leonetti1997} and feedback mediated by voltage-dependent transport properties \cite{Leonetti1998a, Leonetti1998b, Leonetti1999}. In those approaches, bulk ion concentrations and electric potentials evolve dynamically and constitute the primary unstable degrees of freedom.

In contrast, we work in an electroneutral, quasi-static regime appropriate to cellular and tissue scales, where bulk electrodiffusion relaxes rapidly and surface transport activity becomes the relevant dynamical variable. Within this coarse-grained electrohydraulic description, electric and osmotic fields are determined by interfacial currents and act back on the transport activity that generates them. 
Within this unified framework, we compare two distinct feedback mechanisms: (i) transport activity that depends on the electric field normal to the interface, and (ii) transport activity that depends on the electric field tangential to the interface. These feedbacks can arise in biological systems through different physical mechanisms. In single cells, sensitivity to the normal field may reflect voltage-dependent transport, while sensitivity to tangential fields can arise from electrophoretic redistribution of ion transporters \cite{Jaffe1977, Poo1977}. In epithelial tissues, analogous effects may emerge from feedback between trans-epithelial polarity and the normal electric field, or from cell migration in response to tangential electric fields \cite{Shim2024}.  \\

\noindent{\bf Feedback from normal electric field.} 
We first consider the feedback mechanism in which transport activity responds to potential differences across the interface, i.e., to the electric field normal to the interface. We introduce a simplified model in which the activity of an ion pump depends linearly on the local trans-epithelial potential difference. Small fluctuations in pump activity, denoted by $\delta s$, evolve according to:
\be 
\label{eq:dynamics-pump}
\delta \dot s^+ = - \tilde k_n \delta s^+ +  \tilde \alpha_n (\delta\Phi_E - \delta\Phi_I) + D_s \nabla_s^2 \delta s^+.
\ee 
Here, $\tilde k_n$ is a linear relaxation rate, $D_s$ is the surface diffusivity, and the term  $\tilde \alpha_n (\delta\Phi_E - \delta\Phi_I)$ captures the feedback of the radial potential difference on ion pump activity. 

We work in the regime where ion transport dynamics is slow compared to diffusion along the surface. In this limit, the ion concentrations and electric potential profiles rapidly relax to steadystate and are given by Eqs.~\eqref{eq:laplace_c} and~\eqref{eq:laplace_phi} combined with the linearised boundary condition given by Eq.~\eqref{eq:ion-bc}. 

To demonstrate the instability is suffices to consider only $\delta s^+$ and take $\delta s^- =  0$. In the limit $D^+/\Lambda^+ \gg R$ justified above, we obtain the following relations for the spherical harmonic modes of the potential difference across the interface:  
\be 
\beta (\delta \Phi_{lm}^E - \delta \Phi_{lm}^I) = \f{l}{l+1}\f{R \bar \Lambda\delta s^+_{lm}}{ l \bar c_E D^+} + \f{R \bar \Lambda\delta s^+_{lm}}{ l \bar c_I D^+}
\ee 
Substituting this expression into Eq.~\eqref{eq:dynamics-pump} and taking $\bar c_E = \bar c_I$ yields the following evolution equation for the pump activity modes:
\be 
\delta \dot s^+_{lm} = -\f{D_s}{R^2}\l k_n + l(l+1)  - \f{2l+1}{l(l+1)}\alpha_n \r\delta s^+_{lm} ,
\ee 
where $\alpha_n = \tilde \alpha_n R^3 \bar \Lambda/c_I D_s D^+$ and $k_n = \tilde k_n R^2/D_s$.
For large angular modes $l$, the dynamics is always stable due to the dominance of surface diffusion. 
Moreover, since the feedback term proportional to $\alpha_n$ decreases monotonically with increasing $l$ and attains its maximum at $l=1$, stability of the $l=1$ mode guarantees stability of all higher modes, Fig.~\ref{fig4}(B). The system is therefore linearly stable provided $ 2(k_n + 2)  > 3\alpha_n$. This condition defines the parameter regime in which voltage-dependent ion transport destabilizes the homogeneous state and drives spontaneous polarisation. The corresponding phase boundary in the $(k_n,\alpha_n)$ plane is shown in Fig.~\ref{fig4}(C), separating the homogeneous state from the regime of spontaneous electrohydraulic polarisation.\\

\noindent{\bf Feedback from tangential electric fields.}
We next consider a feedback mechanism in which transport activity responds to the electric field tangential to the interface. 
In general, the tangential electric field is not continuous across a permeable interface: the field evaluated just inside the surface differs in sign from that evaluated just outside. Transporter redistribution may therefore be governed either by the interior or exterior tangential field, leading to distinct dynamical consequences.

We model this phenomenologically by assuming that the active transport rate  $\delta s^+$ depends linearly on the local tangential electric field. 
For small fluctuations it evolve according to:
\be 
\label{eq:dynamics-pump-t}
\delta \dot s^+ = - \tilde k_t \delta s^+ +  \tilde \alpha_t \beta D_s \nabla_s^2\delta\Phi_{E,I} + D_s \nabla_s^2 \delta s^+.
\ee 
In the quasi-static electroneutral regime developed above, this yields a closed evolution equation for the spherical harmonic modes $\delta s^+_{lm}$.

If redistribution is governed by the tangential field evaluated on the exterior side of the interface, substituting \refn{eq:sol_phi_main} in \refn{eq:dynamics-pump-t}, the mode dynamics reads:
\begin{equation}
\delta \dot s^+_{lm}
=
-\frac{D_s}{R^2}
\left[
k_t + l(l+1) + l\alpha_t
\right]
\delta s^+_{lm},
\end{equation}
which is linearly stable for all $l$.

In contrast, if redistribution is governed by the tangential field evaluated on the interior side, the mode dynamics, after substituting \refn{eq:sol_phi_main} and \refn{eq:soln-ext} in \refn{eq:dynamics-pump-t} we get
\begin{equation}
\delta \dot s^+_{lm}
=
-\frac{D_s}{R^2}
\left[
k_t + l(l+1) - (l+1)\alpha_t
\right]
\delta s^+_{lm}.
\end{equation}
In this case, sufficiently strong feedback destabilizes a finite-$l$ mode, selecting a characteristic angular scale for symmetry breaking (Fig.~\ref{fig4}(E)). The corresponding stability boundary in feedback–relaxation parameter space is shown in Fig.~\ref{fig4}(F).

Within the same electrohydraulic field theory, instability is generic once feedback is present, but its spatial structure is not: normal-field feedback selects global ($l=1$) polarization, whereas tangential feedback selects finite-$l$ patterned modes. Mode selection is therefore dictated solely by how transport couples to the electric field.  The resulting instability is electrohydraulic in a strict sense: it produces simultaneous electrical, osmotic, and hydrodynamic patterning. Electric potentials, concentration gradients, and fluid flows become spatially organized together, reflecting the inseparability of the coupled fields.

\section{Discussion}

We have developed an electrohydraulic field theory and show that active ion transport at permeable interfaces generates self-consistent electric, osmotic, and hydrodynamic fields across cellular and tissue scales. We derive a general and quantitative mapping between the multipolar structure of interfacial active transport and the bulk electrohydraulic fields. The resulting electrohydraulic fields exhibit substantial osmotic pressure difference across the interface but negligible hydrostatic pressure differences. This is a consequence of the small ratio \(\Lambda_w\eta/R\) in the biologically relevant parameter regime. Thus, anisotropic ion transport can drive fluid flows through osmotic pressure differences without appreciable hydrostatic pressure buildup or tissue deformation.
The estimated field magnitudes are consistent with experimental observations. Electric fields of order $10^{-2}$--$10^{-1}$~V/cm lie within the lower range of endogenous bioelectric fields measured in embryos and regenerating tissues \cite{McCaig2005, Zhao2006}, while the associated osmotic pressures ($1$--$10$~kPa) and flow velocities match those relevant for cellular transport and migration \cite{Trepat2009}. These estimates suggest that electrohydraulic coupling can provide a mechanism for long-range transport and signalling in cells and tissues.

We further study the response of cell and tissues to external electric fields. At linear order, a homogeneous external field induces a dipolar electric field and osmotic pressure gradient. The latter provides a mechanism for self propulsion in an external field without requiring surface charges or electro-osmotic slip. We show that, at linear order, an external electric field generates electrohydraulic fields equivalent to those produced by a dipolar pattern of active ion pumping. Furthermore, a nonlinear electrohydraulic coupling produces an isotropic osmotic pressure  leading to isotropic lumen inflation despite anisotropic forcing. This mechanism provides a natural explanation for electric-field-induced swelling observed in epithelial organoids \cite{Shim2024} and predicts a quadratic scaling of osmotic pressure with field strength.

The theory further reveals a mechanism for spontaneous symmetry breaking through feedback between active ion transport and the electrohydraulic fields it generates. We identify two distinct instabilities, a long-wavelength mode corresponding to global dipolar polarisation, and a finite-wavelength mode associated with lateral pattern formation. These instabilities represent two different routes to symmetry breaking, controlled respectively by normal and tangential electric fields. The results thus establish how different components of electrohydraulic feedback can select qualitatively different spatial patterns.

The observables studied in our theory become rapidly accessible experimentally by new techniques. Recent advances in osmotic \cite{Vian2023} and voltage sensors \cite{Abdelfattah2023}, together with extracellular current mapping techniques \cite{Reid2007}, make it possible in principle to measure all relevant fields within a single system. A decisive test of the theory would require simultaneous mapping of osmolarity, electric potential, and flow fields, allowing direct comparison with the predicted multipolar structure. Importantly, the theory makes falsifiable predictions: (i) a quadratic dependence of isotropic pressure on applied field strength, (ii) self-propulsion velocities controlled by hydraulic permeability rather than surface charge, and (iii) distinct symmetry-breaking modes arising from perturbations of voltage sensitivity versus lateral transporter mobility.

A concrete realisation of this mechanism is provided by any epithelial cavity composed of cells with spatially varying transport properties --- including kidney tubules, lung epithelia, and early embryonic cavities --- that can be treated as an electrohydraulic system. Within the present framework, patterns of ion transport are predicted to produce large-scale electric fields and osmotic pressure gradients which may feedback on tissue mechanics and growth \cite{Saw2022, Mukherjee2026}.

While biological systems can provide a natural realisation, the underlying mechanisms for the generation of electrohydraulic fields is general. Any system with active, spatially heterogeneous transport across a permeable interface --- including synthetic membranes, active emulsions, or engineered tissues --- should exhibit the same electrohydraulic coupling. In this sense, the theory defines a broader class of non-equilibrium systems in which boundary-driven fluxes generate bulk organisation.

In conclusion, our results suggest that osmotic pressure and electric fields should be viewed  as active, self-organised fields that mediate long-range interactions in living and synthetic matter. This perspective opens a route toward a unified description of electrohydrodynamics in active systems, with potential implications for morphogenesis, transport, and pattern formation across a wide range of physical and biological contexts.

The framework can be further extended to account for the porous or gel like nature of the cell interior or surrounding extracellular matrix. In such a description, the fluid flow can differ from that of a Stokes flow; for example, a Darcy flow described by $ {\bf v} \sim \kappa \nabla P $, where $\kappa$ is the hydraulic permeability, could be more appropriate. Multiphase models of cell migration that couple cytosolic water transport, osmotic pressure, ion fluxes, and the cytoskeletal network have been considered previously, for example in Ref.~\cite{Li2018}.

Our simple model based one positive concentration, and one negative concentration does not capture the complexity of biological systems. However, the phenomena that we discuss here are generic and their key features do not depend on the simplifications used. It will be interesting to explore the role of additional ion species in specific biological contexts.

The role of electrohydraulic symmetry breaking in organoid development represents a promising direction for future work. Organoids undergo spontaneous symmetry breaking and spatial organisation during development, raising the possibility that active ion transport and the associated electric and osmotic fields contribute to these processes. Extending the present theory to a dynamically growing tissue and comparing its predictions with spatial measurements of voltage, osmolarity, and fluid flow could provide a direct test of this hypothesis.

\section{Acknowledgment}

We thank Jacques Prost for enlightening discussions. ASV acknowledges funding from Leverhulme Trust (LIP-2021-017). AM acknowledges support from the Max Planck School Matter to Life.

\appendix

\section{Isotropic steady state solution}
We compute the isotropic solutions corresponding to the case of uniformly distributed ion pumps and channels. Spherical symmetry ensures  ($J_r^\pm = 0$) in \refn{eq:ion-bc}, i.e. $\bar c_I =  \bar c_E e^{\mp\beta\Delta \bar \Phi} - \bar s^\pm$, where bar denotes steady state values and $\bar s^\pm = S^\pm/\Lambda^\pm$. 
Solving for $\Delta \bar \Phi$ and $\bar c_I$, we obtain:

\bea 
e^{\beta\Delta \bar \Phi} &=& \sqrt{\frac{1}{4 \bar c_E^2}\l \bar s^+ - \bar s^- \r^2 + 1} - \frac{1}{2 c_E}\l \bar s^+ - \bar  s^- \r ,\quad \\
\bar c_I &=& \sqrt{\frac{1}{4}\l \bar s^+ - \bar s^- \r^2 + \bar c_E^2} - \frac{1}{2}\l \bar s^+ + \bar  s^- \r.
\eea 
The steady state cavity radius $R$ is set by a balance between osmotic and hydrostatic pressures. Using the Young–Laplace relation $\Delta P = 2\gamma /R$ to express the hydrostatic pressure difference in term of the epithelial tension $\gamma$ and the osmotic pressure difference $\Delta \Pi = 2 k_B T (c_I - c_E)$ in \refn{eq:fluid-bc}  determine lumen size. While this solution is unstable for constant $\gamma$, stability can be restored by including elastic contributions to the epithelial tension \cite{Bovyn2024}.

\section{Full solution for anisotropic transport}

We expand the concentration and electrostatic potential perturbations in spherical harmonics as $\delta c^{i} =\sum_{lm}\delta c_{lm}^{i}(r)Y_{lm}$ and $\delta \Phi^{i}=\sum_{lm}\delta \Phi_{lm}^{i}(r)Y_{lm}$, where $i\in{I,E}$.

For $l \neq 0$, substituting $\delta c_{lm}^E = - l \delta c_{lm}^I/(l+1)$ and $\bar c_0\Phi_{lm}^E = - l \bar c_I\Phi_{lm}^I/(l+1)$ in linearized \refn{eq:ion-bc} we get
\bea
\label{eq:ion-bc-lin-modes}
\nn J_{rlm}^\pm &=& \bar\Lambda^\pm \ls \delta c_{lm}^I \l  1 +\f{l}{l+1}  e^{\mp\beta\Delta \bar \Phi} \r \right. \\
 &&\left.  \pm \bar c_E e^{\mp\beta\Delta \Phi} \beta \delta \Phi_{lm}^I\l 1 + \f{l \bar c_I}{(l+1) \bar c_E}\r + s_{lm}^\pm\rs .\qquad
\eea
Substituting this in the flux balance equation \refn{eq:flux-continuity1} we get
\bea
\label{eq:ion-soln}
\nn s_{lm}^\pm + \delta c_{lm}^I \l  1 +\f{l}{l+1}  e^{\mp\beta\Delta \bar \Phi}  +  \f{l D^\pm}{R\bar\Lambda^\pm} \r \pm  &&\\
 \beta \delta \Phi_{lm}^I\l \bar c_E e^{\mp\beta\Delta \bar \Phi} + \f{l \bar c_I e^{\mp\beta\Delta \bar \Phi}}{l+1} + \bar c_I\f{l D^\pm}{R\bar\Lambda^\pm} \r  &=& 0 , \qquad
\eea
From this we get $\delta c_{lm}^I$ and $\delta \Phi_{lm}^I$. We see that the concentration and the electric potential can be varied independently. 
The magnitude of concentration and electric potential depends on the scale of perturbation $\delta s^\pm$, the electric potential difference and the salt concentration in the unperturbed state, the ratio of the two length scale: $D^\pm/\Lambda^\pm$ and $R$, and $l$.

The ion permeability of a tissue can differ by several order of magnitude \cite{Torres2021}. Wide range of $\bar \Lambda = 3 - 3.10^{-3} \mu m/s $ gives $D/R\bar\Lambda^\pm = 10 - 10^4 $, using this the equation simplifies to
\be
\label{eq:ss-modes-approx}
\delta c_{lm}^I \pm  \beta \bar c_I\delta \Phi_{lm}^I  = - \f{R \bar\Lambda^\pm s_{lm}^\pm}{l D^\pm}, 
\ee
which gives
\bea 
\delta c_{lm}^I &=&  - \l \f{R \bar\Lambda^+ s_{lm}^+}{l D^+} +  \f{R \bar\Lambda^- s_{lm}^-}{l D^-}\r,\\
\bar \beta c_I \delta \Phi_{lm}^I &=& - \l \f{R \bar\Lambda^+ s_{lm}^+}{l D^+} -  \f{R \bar\Lambda^- s_{lm}^-}{l D^-}\r.
\eea  
 For $\bar \Lambda^+ = \bar \Lambda^-$ and $D^+ = D^-$ this reduces to Eq.~ \eqref{eq:sol_c_main} and Eq.~ \eqref{eq:sol_phi_main} of the main text.

The fluid flow is obtained by solving the Stokes equations in the interior and exterior regions.
Using the standard Lamb expansion in spherical harmonics \cite{Kim2013} to solve the interior and exterior Stokes equations, together with continuity of velocity and normal-stress balance at the interface, gives
\begin{equation}
\Delta P_{lm}
=
-\frac{\eta}{R}\mathcal K_l\, v^r_{lm}(R),
\label{eq:DeltaP_appendix}
\end{equation}

where the mode-dependent hydrodynamic coefficient is

\begin{equation}
\mathcal K_l
=
\frac{(2l+3)(l+1)^2+l^2(2l-1)}{l(l+1)}
-
\frac{2l-1}{l+1}\delta_{l1}\delta_{m0}.
\label{eq:Kl_appendix}
\end{equation}

The second term accounts for the special treatment of the translational $l=1$ mode.
Combining Eq.~\eqref{eq:DeltaP_appendix} with the permeation boundary condition given by \refn{eq:fluid-bc} of the main text,
yields, mode by mode, the flow field given by \refn{eq:vr_main} in the main text. 

\section{Perturbative solution in external electric field}

The ion concentrations, electric potentials, and surface flux (\refn{eq:ion-bc}), inside ($i = I$) and outside ($i = E$) the cavity are expanded as
\begin{align}
c^{i} &= c_0^i + \epsilon\, c_1^i + \epsilon^2 c_2^i + \mathcal{O}(\epsilon^3),\\
\Phi^{i} &= \Phi_0^i + \epsilon\,(\Phi_1^i + \Phi_{\mathrm{ext}}) 
+ \epsilon^2 \Phi_2^i + \mathcal{O}(\epsilon^3),\\
J_r^\pm &= J_{r,0}^\pm + \epsilon\,J_{r,1}^\pm + \epsilon^2 \,J_{r,2}^\pm +\mathcal{O}(\epsilon^3).
\end{align}

At zeroth order ($\epsilon^0$), the system is isotropic and spatially uniform, i.e., $J_{r,0}^\pm =0$, that gives the following set of equations that solve for zeroth order fields. 
\begin{equation}
\Lambda^\pm \left( c_0^I - c_0^E e^{\mp \beta \Delta \Phi_0} \right) + S^\pm = 0.
\end{equation}

At $\mathcal{O}(\epsilon)$, the flux continuity condition given by \refn{eq:flux-continuity1} reads
\begin{equation}
\partial_r c_1^E \pm \beta c_0^E \partial_r \Phi_1^E 
\mp \beta c_0^E E_0 \delta_{1,l} =
\partial_r c_1^I \pm \beta c_0^I \partial_r \Phi_1^I.
\end{equation}
Adding and subtracting the fluxes for positive and negative ions and expanding the fields in spherical harmonics ($l,m$) yields
\begin{align}
c_{1,lm}^E &= -\frac{l}{l+1} c_{1,lm}^I,\\
(l+1)c_0^E \Phi_{1,lm}^E &= -c_0^E E_0 R\, \delta_{1,l} - l c_0^I \Phi_{1,lm}^I.
\end{align}
Using above relations, at the first order, the boundary flux $J_{r,1}^\pm$ obtained by substituting the fields in \refn{eq:ion-bc} is given by \refn{eq:ion-bc-lin-modes} when the source term is identified as 
\begin{equation}
s_{lm}^{\pm} = 
\pm c_{0}^E e^{\mp \beta \Delta \Phi_0} 
\frac{R \beta q E_0}{2}\, \delta_{1l}\delta_{0m},
\end{equation}
which corresponds to Eq.~\refn{eq:effective-s} of the main text.
The source term is non-zero only for $l=1, m=0$. Therefore, to leading order, only the dipolar mode ($l=1$) is non-zero. 

At $\mathcal{O}(\epsilon^2)$, the bulk equations \refn{eq:laplace_c} and \refn{eq:laplace_phi} of the main text yield
\begin{align}
\nabla^2 c_2^i &= 0,\quad \mathrm{and}\quad
c_0^i \nabla^2 \Phi_2^i = S_i,
\end{align}
where the effective nonlinear term is
\begin{equation}
S_i = \nabla c_1^i \cdot \nabla \Phi_1^i 
- \delta_{i,E}\, q\, \nabla c_1^i \cdot \mathbf{E}_{\mathrm{ext}}.
\end{equation}
Products of the first-order fields generate terms proportional to $Y_{10}^2$ in $S_i$ as well as in the surface flux $J_{r,2}^\pm$. $Y_{10}^2$ written in terms of standard spherical harmonics reads,
\begin{equation}
Y_{10}^2 = \frac{3}{4\pi}\cos^2\theta
= \frac{1}{\sqrt{4\pi}} Y_{00} + \frac{1}{\sqrt{5\pi}} Y_{20}.
\end{equation}
Thus, products of the first-order fields decompose into $l=0$ and $l=2$ modes. This reduces the nonlinear problem to a linear one, where the first order $l=1$ terms drive the second order $l=0$ and $l=2$ modes. We can expand the surface flux as $J_{r,2}^\pm = J_{2,00}^\pm + J_{2,20}^\pm$.

Here we are interested in in the isotropic change in osmotic pressure, that corresponds to $l=0$. Expanding the surface flux to the second order we get
\bea
\nonumber J_{r,00}^\pm = \Lambda^\pm \left( e^{\pm\beta \Delta\Phi_0} c_{2}^I 
 -  c_{2}^E \pm c_{1}^E \beta (\Delta\Phi_1) \right.  \\ 
 \pm  c_{0}^E\beta \Delta \Phi_2 - c_{0}^E \beta^2 (\Delta \Phi_1)^2.
\eea
Setting $J_{r,00}^\pm = 0$ yields two linear equations for $c_2^I$ and $\Delta\Phi_2$, that gives
\be 
 c_{2}^I  = \f{2c_{0}^E (\beta \Delta \Phi_1)^2}{ \l e^{\beta \Delta\Phi_0} + e^{-\beta \Delta\Phi_0}\r} ,
 \label{eq:nonlin-c}
\ee 
where, without loss of generality, we set the exterior isotropic correction $c_2^E=0$. 
Substituting $\Delta \Phi_1$ in  \refn{eq:nonlin-c} gives \refn{eq:pi-E} of the main text.

\bibliographystyle{unsrt}
\bibliography{electrohydraulics}

\end{document}